\documentclass[sigconf]{acmart}

\usepackage{url}
\usepackage{color}
\usepackage{bm}
\usepackage{ulem}
\usepackage{acronym}
\usepackage{comment}
\usepackage{multirow}
\usepackage{graphicx}
\usepackage{subcaption}
\usepackage{siunitx}
\usepackage[capitalise]{cleveref}
\newcommand{\diff}{\,\mathrm{d}}
\newcommand{\numgru}{\num[group-minimum-digits=3]}

\AtBeginDocument{%
  }

\acmConference[]{}{}{}

\acrodef{GP}{genetic programming}
\acrodef{G3P}{grammar-guided genetic programming}
\acrodef{CFG}{context-free grammar}
\acrodef{EA}{evolutionary algorithm}
\acrodef{AI}{artificial intelligence}
\acrodef{AMG}{algebraic multigrid}
\acrodef{GMG}{geometric multigrid}
\acrodef{CGC}{coarse-grid correction}

\renewcommand\footnotetextcopyrightpermission[1]{} 
\begin{document}
\title{Towards Automated Algebraic Multigrid Preconditioner Design Using Genetic Programming for Large-Scale Laser Beam Welding Simulations}
\renewcommand{\shorttitle}{Towards Automated AMG Design Using Genetic Programming for Large-Scale Laser Beam Welding Simulations}

\author{Dinesh Parthasarathy}
\email{dinesh.parthasarathy@fau.de}
\affiliation{%
  \institution{Chair for Computer Science -- System Simulation, Friedrich-Alexander-Universität
Erlangen-Nürnberg}
  \city{Erlangen}
  \country{Germany}
}
\author{Tommaso Bevilacqua}
\email{tommaso.bevilacqua@uni-koeln.de}
\affiliation{%
  \institution{Department Mathematik/Informatik, Universit{\"a}t zu K{\"o}ln}
  \city{K{\"o}ln}
  \country{Germany}
}

\author{Martin Lanser}
\email{martin.lanser@uni-koeln.de}
\affiliation{%
  \institution{Department Mathematik/Informatik and Center for Data and Simulation Science, Universit{\"a}t zu K{\"o}ln}
  \city{K{\"o}ln}
  \country{Germany}
}

\author{Axel Klawonn}
\email{axel.klawonn@uni-koeln.de}
\affiliation{%
  \institution{Department Mathematik/Informatik and Center for Data and Simulation Science, Universit{\"a}t zu K{\"o}ln}
  \city{K{\"o}ln}
  \country{Germany}
}
\author{Harald Köstler}
\email{harald.koestler@fau.de}
\affiliation{%
  \institution{Chair for Computer Science -- System Simulation, Friedrich-Alexander-Universität
Erlangen-Nürnberg}
  \city{Erlangen}
  \country{Germany}
}

\renewcommand{\shortauthors}{Parthasarathy et al.}

\begin{abstract}
Multigrid methods are asymptotically optimal algorithms ideal for large-scale simulations. But, they require making numerous algorithmic choices that significantly influence their efficiency. Unlike recent approaches that learn optimal multigrid components using machine learning techniques, we adopt a complementary strategy here, employing evolutionary algorithms to construct efficient multigrid cycles from available individual components.

This technology is applied to finite element simulations of the laser beam welding process. The thermo-elastic behavior is described by a coupled system of time-dependent thermo-elasticity equations, leading to nonlinear and ill-conditioned systems. The nonlinearity is addressed using Newton’s method, and iterative solvers are accelerated with an \ac{AMG} preconditioner using \textit{hypre} BoomerAMG interfaced via PETSc. This is applied as a monolithic solver for the coupled equations.

To further enhance solver efficiency, flexible \ac{AMG} cycles are introduced, extending traditional cycle types with level-specific smoothing sequences and non-recursive cycling patterns. These are automatically generated using genetic programming, guided by a context-free grammar containing \ac{AMG} rules.  Numerical experiments demonstrate the potential of these approaches to improve solver performance in large-scale laser beam welding simulations.

\end{abstract}

\ccsdesc[500]{Computing methodologies~Genetic programming}
\ccsdesc[300]{Computing methodologies~Massively parallel and high-performance simulations}
\ccsdesc[500]{Mathematics of computing~Solvers}

\keywords{Algebraic Multigrid, Artificial Intelligence, Evolutionary Algorithms, Genetic Programming, Laser Beam Welding, Nonlinear Finite Element Method, Thermo-elasticity}


\maketitle

\section{Introduction}
Laser beam welding is a non-contact method for joining materials that has gained significant importance as industrial production becomes more automated. Its appeal lies in the short cycle times and minimal heat-affected zones it offers. However, the rapid cooling rates characteristic for this process can result in a residual melt oversaturated with specific alloy elements, potentially causing solidification cracks. Gaining a quantitative understanding of how these fractures develop and their relationship to process parameters is crucial for refining and enhancing welding processes. This specific focus is central to the efforts of the DFG research group 5134 ``Solidification Cracks During Laser Beam Welding -- High Performance Computing for High Performance Processing'' \footnote[3]{https://www.for5134.science/en/}.
Another central part of the research group is to use the potential of modern supercomputers and high performance simulations to get further insights into the process of laser beam welding. A crucial part thereof is the optimization of parallel efficient iterative solvers for large thermo-mechanical finite element simulations. Especially, the latter aspect is the focus of this article.

Here, the thermo-mechanical behavior of laser beam welding is described by a coupled set of time-dependent thermo-elasticity equations, as outlined in \cite{simo1992associative}. Due to the complexity of the problem, the linear systems resulting from finite element discretization in space, backward Euler discretization in time, and linearization with Newton’s method tend to be ill-conditioned. Besides some work on developing and implementing parallel monolithic domain decomposition preconditioners for thermo-mechanical finite element problems~\cite{bevilacqua2024monolithic}, we are also interested in finding efficient \acf{AMG} preconditioners for such problems. As a result, in this work, we explore an automated approach for the design of \ac{AMG} methods. More specifically, the scope of this work is to find optimal configurations for \textit{hypre} BoomerAMG \cite{HENSON2002155, 10.1007/3-540-31619-1_8} as a preconditioner and accelerate the convergence of the GMRES (Generalized Minimal Residual) method; in our numerical simulation we use our software package FE2TI~\cite{klawonn2020computational} based on PETSc \cite{balay2019petsc} with an interface to FEAP \cite{taylor2014feap}.
Here, we apply BoomerAMG directly to the coupled problems arising from the linearization. Unlike approaches that exploit special structures, such as -- defining block-triangular preconditioners and applying \ac{AMG} on separate blocks -- the goal here is to find an optimal BoomerAMG configuration that delivers the best performance when used as a monolithic preconditioner for coupled thermo-elasticity problems.

Multigrid methods have several algorithmic components, such as, smoothers, cycle types, intergrid operators, and coarsening schemes, all of which determine their efficiency. These choices are not only complex but also problem-dependent. Leveraging recent advances in artificial intelligence (AI), efforts have been made to find optimal multigrid components, such as neural smoothers \cite{Huang2023-jb,Zhang2024}, intergrid operators \cite{pmlr-v97-greenfeld19a,10.5555/3524938.3525540, Katrutsa2020, katrutsa2017deep}, and coarsening schemes \cite{Taghibakhshi2021OptimizationBasedAM}. But unlike previous efforts, we take a complementary approach towards constructing efficient multigrid cycles from a set of available multigrid components. Traditional multigrid methods employ recursive cycle types such as V-, W-, and F-cycles. However, we consider an extended search space of multigrid cycles—referred to as \textit{flexible} cycles—with level-specific smoothing sequences and arbitrary non-recursive cycling patterns across the grid hierarchy. The aim is to exploit this \textit{flexibility} and design more efficient multigrid methods than the standard types.
Due to the prohibitively large search space, manually designing a flexible cycle is impractical, so it is formulated as a program synthesis task. Previous work by Schmitt et al. generated such flexible geometric multigrid programs using grammar-guided genetic programming (G3P)  \cite{Schmitt2021, Schmitt2022}. This approach has been extended to generate efficient flexible algebraic multigrid (AMG) methods \cite{ParthasarathyCM2024}. We intend to integrate the latter into the laser beam welding simulation software for automated \ac{AMG} preconditioner design. As a preliminary step, this paper explores the potential benefits of such flexible \ac{AMG} preconditioners, generated offline, decoupled from the simulation, using G3P.\newline
The remainder of the paper is organized as follows: In \cref{sec:thermo-eq} we present the thermo-elasticity equations and the problem of laser beam welding, in \cref{sec:FEMdisc} we introduce the discretization in space and time, in \cref{sec:back} we discuss key concepts of \ac{GP}, in \cref{sec:prec} we describe our approach towards automated AMG design, and finally in \cref{sec:Numres} we show some numerical results.

\begin{figure}[b]
  \centering
  \includegraphics[width=0.5\linewidth]{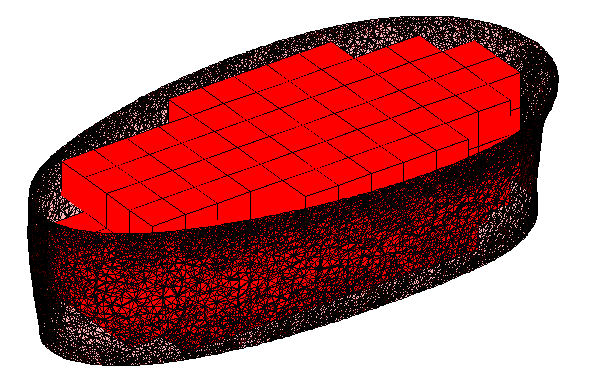}
  \caption{Triangulated surface representing the geometry of the melting pool \cite{bakir2018numerical} and an example of a discrete representation of it using hexahedral elements.}\label{fig:meltpool}
\end{figure}

\section{Laser Beam welding problem}\label{sec:thermo-eq}
{We start presenting the general framework of our problem.} To focus primarily on evaluating the performance of the preconditioners, we streamline the model problem by simplifying the laser beam's action. Consequently, we use experimental data to determine the geometry of the melting pool, that is, the region where the material is melted by the laser. Within this region, we impose the melting temperature of the material across all degrees of freedom (DOFs) by applying Dirichlet boundary conditions. \newline
It is important to note that both the laser beam and the associated melting pool geometry, including the relevant Dirichlet boundaries, change position over time, moving with a problem-specific speed. The interested reader can find a more realistic laser heat source model which has been introduced to achieve more realistic simulations in~\cite{hartwig2024physically, hartwig2023numerical, hartwig2023volumetric}. For context, \cref{fig:tempfiled} offers a general outline of a typical model geometry undergoing laser beam welding, while \cref{fig:meltpool} illustrates the specific geometry of the melting pool used in this work. \newline 

\subsection{Thermo-elasticity equations}
The thermo-mechanical behavior of a plate of metal welded by a laser is described by the set of simplified thermo-elasticity equations
\begin{align*}
 {\rm div} ({\sigma(\bm{u},\theta)}) &= 0, \\
 {\rm div} (\bm{q})+\gamma \, {\rm div} (\dot{\bm{u}}) \, \theta +c_\rho \, \dot{\theta} &= 0,
\end{align*}
where $\bm{u}(x,t)$ and $\theta(x,t)$ represent the displacement vector and the absolute temperature respectively. A complete explanation of the process that leads to this simplified model can be found in \cite{bevilacqua2024monolithic}. Let us remark that for realistic simulations a consideration of the plastic behavior of the material is necessary, that is, the use of thermo-elastoplastic material models. Here, for the initial investigations on optimizing \ac{AMG}, we restrict ourselves to this simplified elastic model.\newline
The material-specific and temperature-dependent material parameters are: $c_\rho$ the heat capacity, $\bm{q}$ the heat flux vector, and $\gamma = 3\alpha_T\kappa$ the stress temperature modulus. Moreover, $\alpha_T$ denotes the coefficient of linear thermal expansion, $\kappa = E / 3(1-2\nu)$ the bulk modulus and $E$ the Young's modulus. Here, we choose the temperature-dependent parameters to model an austenitic chrome-nickel steel(1.4301) as shown in \cref{tab:mat_param}. 
The stress in tensor notation is defined as
\begin{equation*}
\bm{\sigma} := \mathbb{C} : \varepsilon- 3\alpha_T \kappa (\theta -\theta_0) \bm{1}
\end{equation*}
with $\mathbb{C} = \kappa \bm{1} \otimes \bm{1} + 2\mu \mathbb{P}$ and the fourth-order deviatoric projection tensor $\mathbb{P} = \mathbb{I} - \frac{1}{3} \bm{1} \otimes \bm{1}$ and $\theta_0$ the initial temperature of the metal. \\
 The heat flux vector is defined by Fourier's law as $\bm{q} = -\bm{\lambda} \cdot \nabla \theta$, where $\bm{\lambda}$ represents the identity matrix scaled by the thermal conductivity coefficient. As already stated above, the sole source of heat is a laser, modeled through a volumetric Dirichlet boundary condition to represent the molten metal pool that moves across the plate. To obtain a realistic scenario, the geometry of the melting pool is derived from experimental data, resulting in a triangulated surface, as described in \cite{bakir2018numerical} and as shown in \cref{fig:meltpool}.
 
\begin{table*}[t]
     \caption{Parameters of the material austenitic chrome-nickel steel(1.4301) at different temperatures. The values, taken from~\cite{hartwig2023volumetric}, have been provided by Bundesanstalt f{\" u}r Materialforschung und -pr{\" u}fung (BAM).  A linear interpolation is used between the temperature intervals.}\label{tab:mat_param}
  \begin{tabular}{lrrrrrrrrr}
    \toprule
    \multirow{2}{*}{Parameter} & \multicolumn{9}{c}{Value} \\
                         & \multicolumn{9}{c}{(Temperature $[C]$)}\\
    \midrule
    \multirow{2}{*}{$E \cdot 1e4 \, [{\rm N/mm}^2]$} & 20.0 & 19.1 & 17.5 & 12.5 & 7.2 & 1.6 & 0.1 & -- & --\\
     & (20) & (170) & (400) & (800) & (1000) & (1100) & (1500) & -- & -- \\[1mm]     
    \multirow{2}{*}{$\nu $} & 0.271 & 0.284 & 0.300 & 0.319 & 0.329 & 0.364 & 0.364 & -- & -- \\ 
     & (20) & (183) & (484) & (799) & (994) & (1994) & (2000) & -- & -- \\[1mm]
    \multirow{2}{*}{$\alpha_T \cdot 1e-5 \, [{\rm K}^{-1}]$} & 1.6 (20) & 1.81 & 1.98 & 2.13 & 2.23 & 2.23 & 2.33 & -- & -- \\ 
    & (20) & (200) & (580) & (1000) & (1200) & (1500) & (2000) & -- & -- \\[1mm]
    \multirow{2}{*}{$\bm{\lambda} [{\rm W}/({\rm m}\cdot {\rm K}))]$} & 15.6 & 18.1  & 21.0 & 23.8 & 26.6  & 34.4 & 35.0 & 60 & -- \\
    & (20) & (200) & (400) & (600) & (800) & (1350) & (1393) & (1460) & -- \\    
    \multirow{2}{*}{$c_{\rho}  \cdot 1e5 \, [{\rm J/kg}\cdot {\rm K}]$ }& 5.11 & 5.42 & 5.75 & 6.05 & 6.30 & 6.85 & 7.30 & 20.20 & 50.00 \\[1mm]
    & (20) & (200) & (400) & (600) & (800) & (1350) & (1427) & (1442) & (1460) \\
  \bottomrule
\end{tabular}
\end{table*}

The weak formulation of these equations is given by
\begin{align*}
\int_\Omega \sigma (\bm{u})\, :\, \varepsilon (\bm{v}) \diff x &= 0 \,\, \forall \bm{v} \in [H^1(\Omega)]^3, \\
\int_\Omega \bm{q} \cdot \nabla q \diff x + \int_\Omega \{- 3\alpha_T \, \kappa \,{\rm tr[\dot{\varepsilon}] \, \theta - \text{c}_\rho \dot{\theta}} \}q \diff x &= 0 \quad \,\forall q \in H^1(\Omega),
\end{align*}
with Dirichlet boundary conditions $\bm{u} = 0$ on $\partial \Omega_{D,u} \subset \partial \Omega$ and $\theta = \theta_{l} \gg~\!\!0$ on $\partial \Omega_{D,\theta} \subset \partial \Omega$, where $\theta_l$ represents the temperature inside of the melting pool described by $\partial \Omega_{D,\theta}$. This is a nonlinear and nonsymmetric saddle point system, for which the theoretical framework is still an ongoing research topic, such as the investigation of pairs of inf-sup-stable finite elements. However, a theoretical analysis of this problem is beyond the scope of this work.

Since we aim for large and highly resolved simulations, an iterative and parallel solution of the resulting linearized systems is necessary.
Due to their properties, these specific linear systems require appropriate preconditioners to enhance the convergence rate of the GMRES method.

The following section briefly describes the finite element discretization in space, the time discretization, and the linearization with Newton's method. Still, the focus here lies on the efficient solution of the linearized systems, and for a fast read, one can simply skip to \cref{sec:back}.

\section{Finite element discretization}\label{sec:FEMdisc} 
Applying the Newton-Raphson method to the nonlinear boundary value problem results in a linear problem for each nonlinear iteration; further details on this process can be found in \cite{hartwig2024physically}. This problem is spatially discretized using a mixed finite element method. 

Let then $\tau_h$ be a uniform mesh of $N_e$ hexahedral elements $\Omega_e$ of $\Omega$ with characteristic mesh size $h$. We introduce the conforming discrete piecewise linear displacement and temperature spaces
\begin{align*}
V^h &= V^h(\Omega) = \{\bm{u}\in [\mathcal{C}^0(\Bar{\Omega}) \cap H^1(\Omega)]^3: \bm{u}|_T \in Q_1 \ \forall T \in \tau_h\}, \\
Q^h &= Q^h(\Omega) = \{\theta \in \mathcal{C}^0(\Bar{\Omega}) \cap H^1(\Omega): \theta|_T \in Q_1 \ \forall T \in \tau_h\}
\end{align*}
respectively, of Q1-Q1 mixed finite elements. Here, $\mathcal{C}^0(\Bar{\Omega})$ denotes the space of continuous functions on $\Bar{\Omega}$ and $H^1(\Omega)$ the usual Sobolev space. Due to the lack of a stable theoretical framework, we cannot guarantee that the selected finite elements satisfy the inf-sup condition. Time discretization is performed using the backward Euler method with a time step size of $\Delta t$. Consequently, in each time step, we have to solve a linearized system until the global absolute residual from Newton's method is below a chosen tolerance. The resulting discrete saddle point problem for each Newton iteration in each time step finally takes the form
\begin{equation}\label{linsyst}
    K\varDelta \bm{d} = \begin{bmatrix}
        K_{uu} & K_{u\theta} \\
        K_{\theta u} & K_{\theta \theta}
    \end{bmatrix} \begin{bmatrix}
        \varDelta \bm{d}_u \\ \varDelta \bm{d}_\theta 
    \end{bmatrix} = \begin{bmatrix}
        R_u \\ R_\theta 
    \end{bmatrix} = R, 
\end{equation}
where $\varDelta \bm{d}_u$ and $\varDelta \bm{d}_\theta$ represent the Newton update for the displacement and temperature, and $R_u$ and $R_\theta$ the vectors of the residual, respectively. The block matrices are obtained by a standard finite element assembly and we obtain
\begin{equation*}
\begin{split}
& K_{uu} = \mathbb{A}^{N_e}_{e=1} \bigg[ \int_{\Omega_e} \bm{B}_u^T \, \mathbb{C}\, \bm{B}_u \diff x \bigg],\\
& K_{u\theta} = \mathbb{A}^{N_e}_{e=1} \bigg[ -\int_{\Omega_e} \bm{B}_u^T \, (3\alpha_T \kappa \bm{1}^T)\, {\rm \bm{N}}_\theta \diff x \bigg],\\
& K_{\theta u} = \mathbb{A}^{N_e}_{e=1} \bigg[-\frac{1}{\varDelta t} \int_{\Omega_e} \theta \, {\rm \bm{N}}_\theta^T \, (3\alpha_T \kappa \bm{1}) \, \bm{B}_u \diff x \bigg], \\ 
& K_{\theta\theta} = \mathbb{A}^{N_e}_{e=1} \bigg[ -\int_{\Omega_e} \bm{B}_\theta^T \, \bm{\lambda} \, \bm{B}_\theta \diff x \, - \int_{\Omega_e} {\rm \bm{N}}_\theta^T \, {\rm tr}[\dot{\varepsilon}] \, {\rm \bm{N}}_\theta \diff x \\
& \qquad \qquad  \,- \frac{1}{\varDelta t}\int_{\Omega_e} {\rm \bm{N}}_\theta^T \, c_\rho \, {\rm \bm{N}}_\theta \diff x\bigg],
\end{split}
\end{equation*}
where $\bm{N}_u$ and ${\rm \bm{N}}_\theta$ are the finite element nodal basis functions for displacement and temperature, and $\bm{B}_u$ and $\bm{B}_\theta$ denote their derivatives in tensor notation. 


\section{Genetic Programming: A Primer}\label{sec:back}
Before delving into the automated design of \ac{AMG} methods, we briefly introduce the key concepts of \ac{GP}. \ac{GP} is an \ac{AI} technique that uses a metaheuristic approach for the evolution of computer code. It is a branch of \acp{EA}, rooted in Darwin's principle of natural selection. The general approach of GP is to apply a population of programs to a given problem, and compare their performance (fitness) relative to each other. Operators inspired by genetics (crossover, mutation) are applied to selected programs from the population (parents) such that in time better programs (offspring) emerge by evolution. This principle is applied iteratively for multiple generations until a desirable population of programs is obtained \cite{Banzhaf2001, Orlov2009}. 
In problems where the programs are required to conform to a specific structure (for eg. \ac{AMG} preconditioners), grammar rules can be used to constrain the evolutionary process such that only valid programs are produced. \Acf{G3P} is an extension of traditional GP systems that use \acp{CFG} to impose such constraints on the initial population and subsequent genetic operations. Exploiting domain knowledge using \acp{CFG} helps in eliminating invalid programs and speeds up the optimization process \cite{Whigham1995GrammaticallybasedGP, Manrique2009}.

\begin{figure}
    \centering
    \includegraphics[width=1\linewidth]{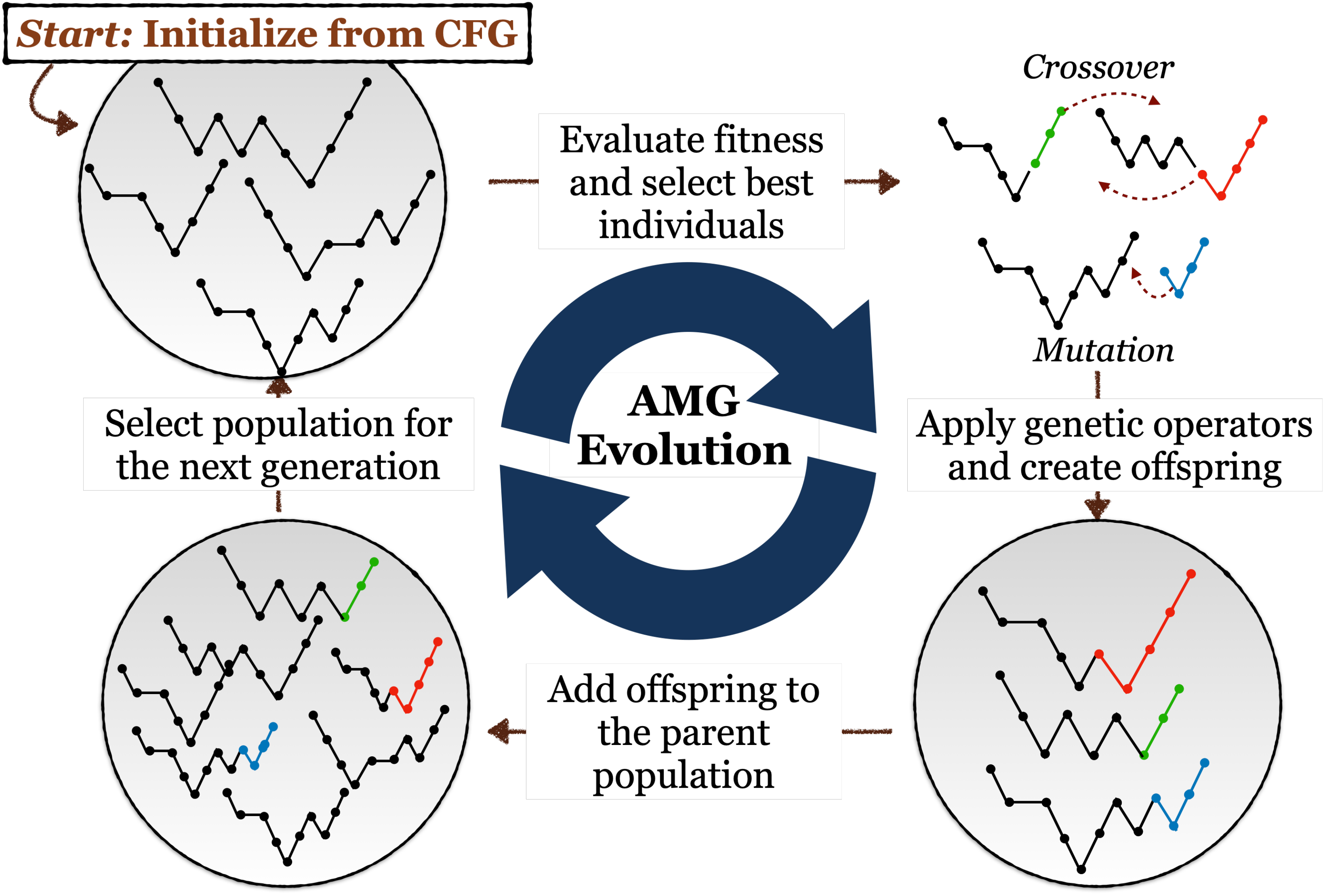}
    \caption{A simplified illustration of the evolution of \ac{AMG} cycles using G3P.}
    \label{fig:GPoverview}
\end{figure}

\ac{G3P} for evolving \ac{AMG} cycles is illustrated in \cref{fig:GPoverview}. The \ac{CFG} generates an initial random population of \textit{flexible} cycles. The individuals in the population are evaluated for their \textit{fitness}, for example, solve time and convergence rate, and the best ones are selected. Crossover and mutation operators are applied to these individuals, constrained by the \ac{CFG}. During crossover, parts of the \ac{AMG} cycle between pairs of selected individuals are exchanged (for example, red and green portions in \cref{fig:GPoverview}), and during mutation, random perturbations are introduced (for example, the blue portion in \cref{fig:GPoverview}) to encourage exploration. The \ac{CFG} ensures that the crossover points and perturbations are such that the offspring produced are still valid \ac{AMG} programs. The offspring are combined with the parent population, from which the best individuals form the population for the next generation. Repeating these steps iteratively produces a population of efficient \ac{AMG} programs with optimal fitness measures.

\begin{table}[t]
  \caption{Timings for a 1.0\,s simulation with the default BoomerAMG preconditioner setup in PETSc for different problem sizes: 89\,100, 347\,116 and 1\,370\,028 DOFs, with 8, 32 and 128 MPI processes. The timings refer to: assemble the linear systems (\textit{T\textsubscript{ass}}), assemble the \ac{AMG} preconditioners (\textit{T\textsubscript{setUp}}), solve the linear systems (\textit{T\textsubscript{sol}}) and total time for the simulation (\textit{T\textsubscript{tot}})}.
  \label{tab:timings_amg_ref}
  \begin{tabular}{rrrrr}
    \toprule
    DOFs &  $T_{ass}$ & $T_{setUp}$ & $T_{sol}$ & $T_{tot}$\\
    \midrule
    \numgru{89100}   & 6.6\,s & 1.8\,s &  22.7\,s & 31.1\,s \\
    \numgru{347116}  & 7.4\,s & 3.4\,s &  31.9\,s & 42.7\,s \\
    \numgru{1370028} & 8.9\,s & 6.8\,s & 436.7\,s & 452.4\,s \\
  \bottomrule
\end{tabular}
\end{table}

\begin{figure}[b]
    \centering
    \includegraphics[width=1.1\linewidth]{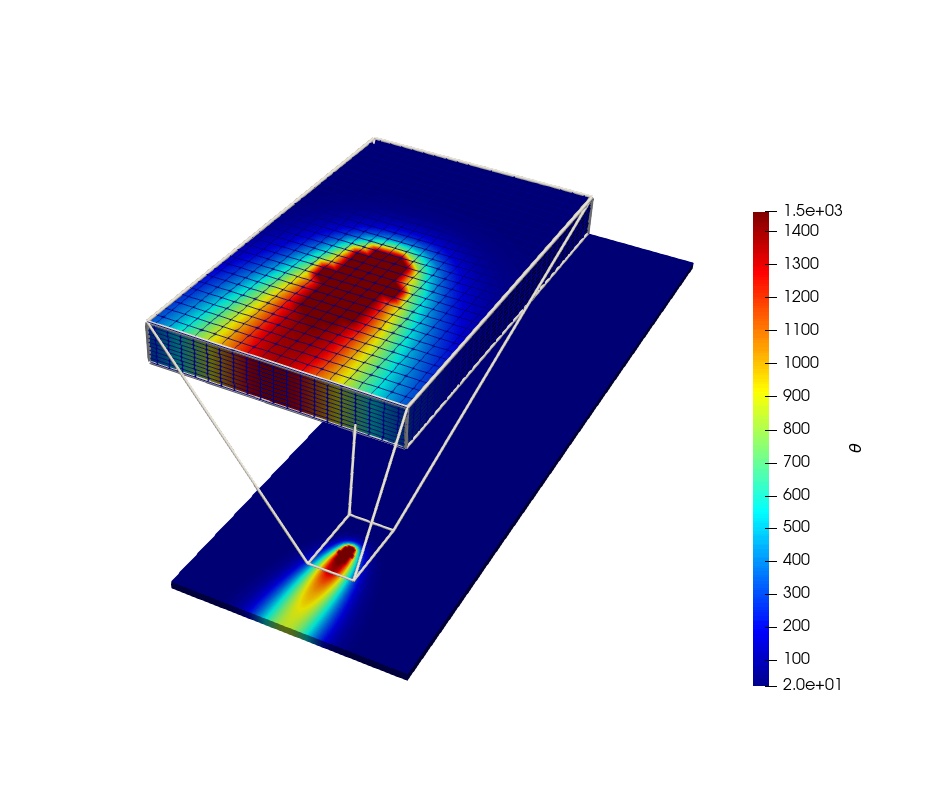}
    \caption{Representation of the temperature field of a fully coupled thermo-elasticity laser beam welding simulation at time 1.0 s, computed on 128 MPI processes with 1.37 million DOFs.} \label{fig:tempfiled}
\end{figure}

\section{Automated \ac{AMG} design}\label{sec:prec}
To solve the linearized systems in our numerical simulations, we use \ac{AMG} as a preconditioner to accelerate the convergence of GMRES, employing \textit{BoomerAMG} from the \textit{hypre} library. In particular, \ac{AMG} is applied with a monolithic approach on the entire coupled displacement-temperature system. Formulating an efficient \ac{AMG} method is non-trivial, involving complex design choices for the \textit{setup phase} (coarsening strategy, interpolation scheme, threshold parameters, etc.) and the \textit{solve phase} (cycle type, smoother, relaxation weights, etc.). In this paper, we focus on the latter, as the \textit{solve phase} is shown to dominate the cost of our simulations (see \cref{tab:timings_amg_ref}). We fix the setup parameters (see \cref{tab:amgsetupparam}) and automate the design of the solve phase, that is, generating efficient \ac{AMG} cycles.

The \ac{AMG} methods are constructed here with cycles that go beyond the standard V-, F-, or W-cycle types. These so-called \textit{flexible cycles} exhibit the following attributes:

\begin{itemize}
    \item the different grid levels can be traversed arbitrarily and non-recursively;
    \item at each grid visitation, the smoothing sequences -- smoother types, number of sweeps, and weights -- can be chosen independently of those at other grid levels or during previous visitations at the same level.
\end{itemize}
Such \textit{flexible} \ac{AMG} methods are automatically generated using a formulated context-free grammar (CFG). The CFG generates an initial population of random \textit{flexible} \ac{AMG} preconditioners and further enforces constraints on the genetic operators—crossover and mutation. In this way, we infuse solver knowledge into the evolutionary algorithm, ensuring all programs in the population conform to valid instances of \ac{AMG}-preconditioned GMRES solvers. The CFG is integrated into the  \textit{EvoStencils} framework\footnote{a software library for the automatic design of multigrid methods} and coupled with an extended implementation of BoomerAMG that supports non-standard flexible cycles \cite{ParthasarathyCM2024}. Refer to \cref{subsec:software} for details on the software pipeline. The population of \textit{flexible} BoomerAMG programs, generated by the CFG and evolved through genetic operators are evaluated for two objectives -- \textit{solve time per iteration} and \textit{convergence}. These fitness measures guide a multi-objective search with \ac{G3P}. Running this optimization for a fixed number of generations finally yields a population of Pareto optimal \ac{AMG}-preconditioned GMRES programs (see \cref{fig:paretofront}).
\subsection{Problem setup}\label{NumRes}
In our numerical simulation we try to reproduce a physically realistic process of laser beam welding. To do so, we solve a thermomechanical problem on a domain $\Omega$ = $100 \, {\rm mm} \times 30\, {\rm mm} \times 1 \,{\rm mm}$, {where} $(0,0,0)$ is one of the corners. The mechanical constraints are enforced by imposing  Dirichlet boundary conditions for the displacements on the faces $y = 0\,{\rm mm}$ and $y = 30\,{\rm mm}$. \\
Our configuration models a laser moving along a plate of an austenitic chrome nickel steel, whose material parameters vary depending on the temperature; see \cref{tab:mat_param}. 
The action of the laser is implemented as a volume Dirichlet boundary condition which represents the boundary of the region of the metal that has been melted (melting pool) as described before; see \cref{fig:meltpool}. Therefore, we load a surface model obtained from experimental data that models the region of the melting pool and we move this geometry along the $x$-direction during the time iterations with a velocity of $1.0 \, m/s$ (\cref{fig:tempfiled}). The initial temperature of the metal is set to $\theta_0 = 20^\circ $C everywhere and the temperature inside of the melting pool is set, as a gradient, to $\theta_l = 1460^\circ $C.
Since the thermo-elastic model does not require particular restrictions on the time discretization, we simulate here $T = 1.0 \, s$ of the laser beam welding process using a backward Euler method with step size $\Delta t = 0.1 \, s$. In each time step Newton's method is applied until the global residual is below an absolute tolerance of $1e-4$, while for the linear systems, a relative residual of $1e-4$ or an absolute residual of $1e-8$ of the unpreconditioned norm is chosen as stopping criterion for the GMRES method. \\

\begin{figure}[t]
    \centering
    \includegraphics[width=1.1\linewidth]{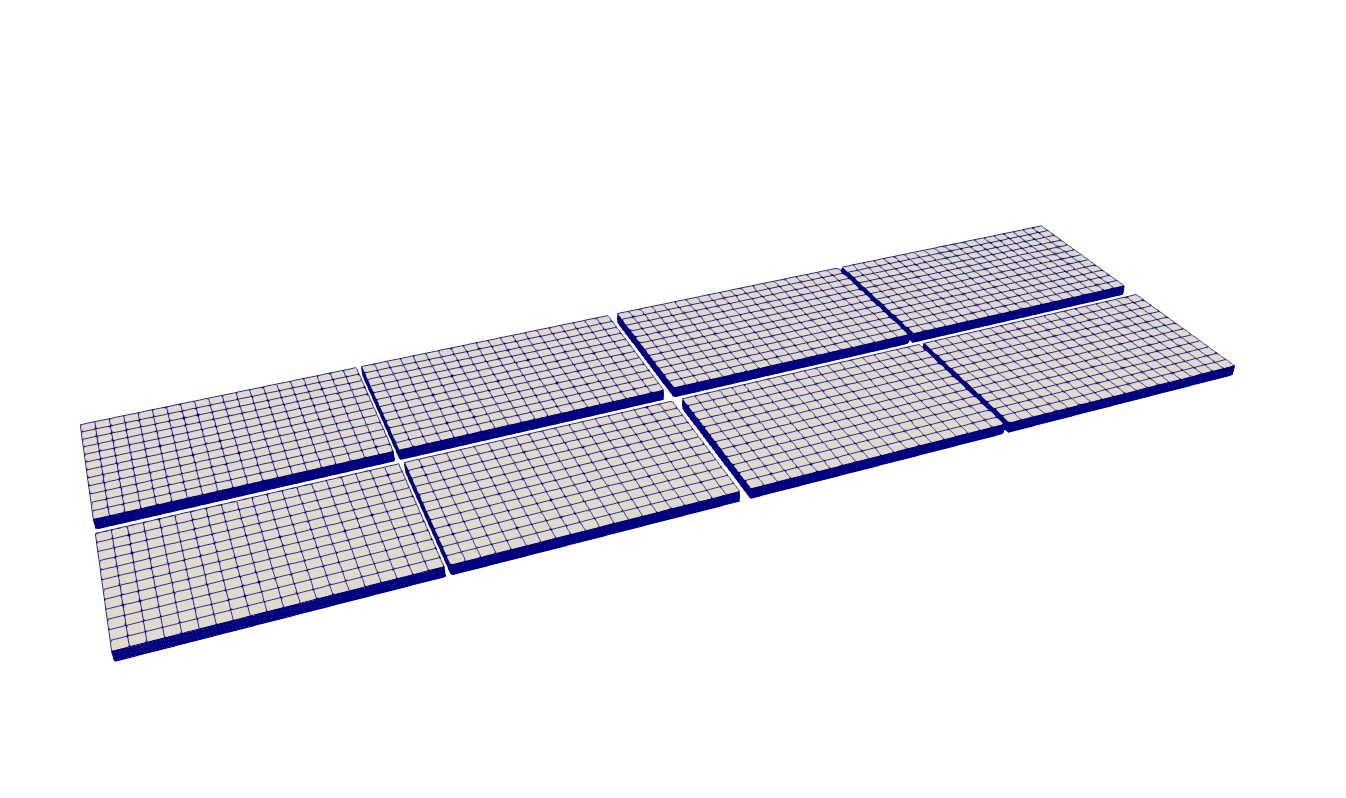}
    \caption{Processor topology for MPI parallelism with 89\,100 DOFs on 8 processes, where each tile of the welding plate is assigned to a separate process.} \label{fig:MPIdist}
\end{figure}

\subsection{Optimization setup}
To run our optimization, we consider the linear system at $t=0.4\,s$  and the $2^{nd}$ Newton iteration as a proxy for the laser beam welding simulation. The \ac{AMG}-preconditioned GMRES solver is designed based on fitness measures (solve time per iteration, convergence) evaluated for this linear system, with \numgru{89100} DOFs distributed across 8 MPI processes. Different parts of the domain are assigned to each MPI process, such that, all elements and their associated DOFs in a subdomain are allocated to the same process (\cref{fig:MPIdist}). This reduces MPI communication, preserves sparsity patterns, and enables a more efficient coarsening.
We generate \ac{AMG} cycles with a \textit{flexible part} and a \textit{recursive part}.  The top fine grid levels ($N_{flex}$) employ a \textit{flexible structure}, while the rest of the coarser levels use the standard default BoomerAMG configuration in \textit{hypre}. In this way, we limit our search space to the most expensive levels, and also retain the possibility to adapt the generated solvers to different problem sizes, by adjusting the height of the \textit{recursive part} (\cref{subsec:weakscaling}). 

The \textit{flexible part} is generated from a set of \ac{AMG} components listed in \cref{tab:amgcomponents}. We consider various relaxation schemes, relaxation ordering, discretely sampled relaxation weights for local relaxation within each processor ($\omega_i$) and Jacobi exchange across processor boundaries ($\omega_o$) \cite{Yang04,Baker2011}, and a set of scaling factors ($\alpha$) for the coarse-grid correction (CGC) step to construct optimal flexible cycles. Details of the GP setup are provided in \cref{subsec:software}.

\begin{table}[h]
  \centering
  \begin{tabular}{|r|l|}
    \hline
    \textbf{Smoothers} & GS-Fwd., GS-Bwd., Jacobi \\ 
                       & $l_1$ GS-Fwd., $l_1$ GS-Bwd., $l_1$ Jacobi \\ 
    \textbf{Relaxation Ordering} & Lexicographic, CF\\ 
    \textbf{Inner Relaxation} ($\omega_i$) & (0.1, 0.15, 0.2, ..., 1.9)\\ 
    \textbf{Outer Relaxation} ($\omega_o$) & (0.1, 0.15, 0.2, ..., 1.9)\\ 
    \textbf{CGC Scaling} ($\alpha$) & (0.1, 0.15, 0.2, ..., 1.9)\\ 
    \textbf{Num. flex. levels} ($N_{flex}$) & 5 \\ 
    \textbf{Num. precond. cycles} & 1 \\ 
    \textbf{Coarse-grid solver} & Gaussian Elimination \\ \hline
  \end{tabular}
  \caption{AMG components for flexible cycle generation.}
  \label{tab:amgcomponents}
\end{table}
\section{Results}\label{sec:Numres}
\begin{figure}[b]
    \centering
    \begin{tabular}{|@{}c@{}|@{}c@{}|}
    \toprule
      \includegraphics[width=0.235\textwidth]{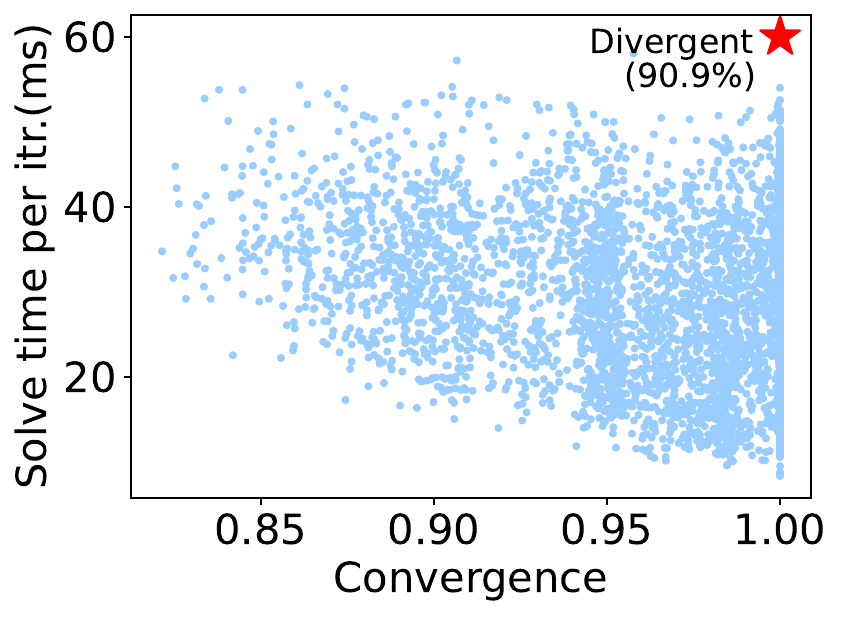} &
      \includegraphics[width=0.235\textwidth]{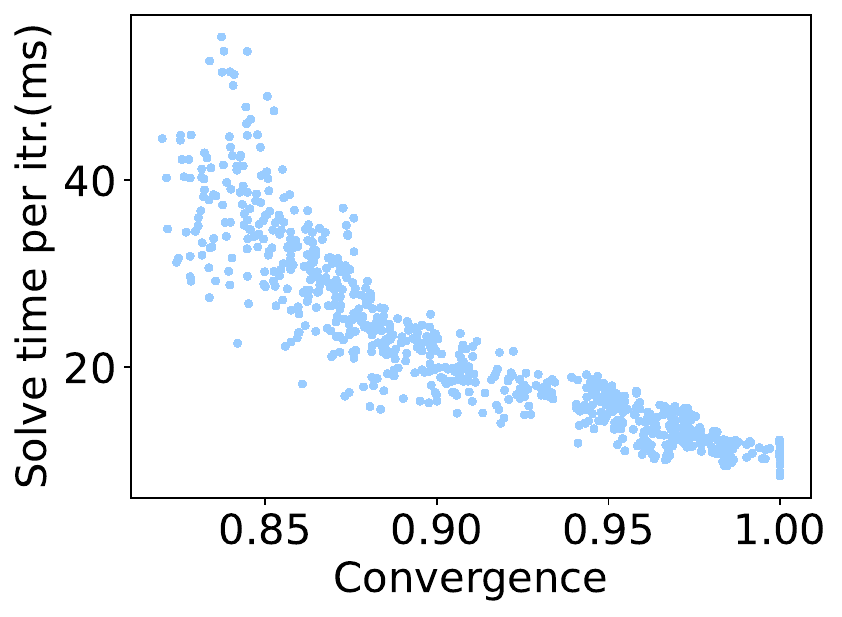} \\
      \midrule
      \includegraphics[width=0.235\textwidth]{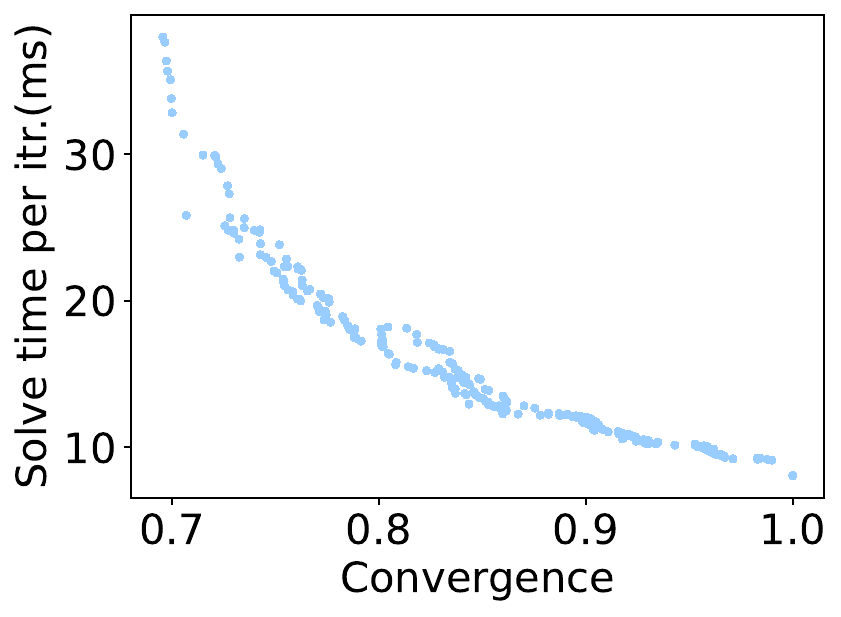} &
      \includegraphics[width=0.235\textwidth]{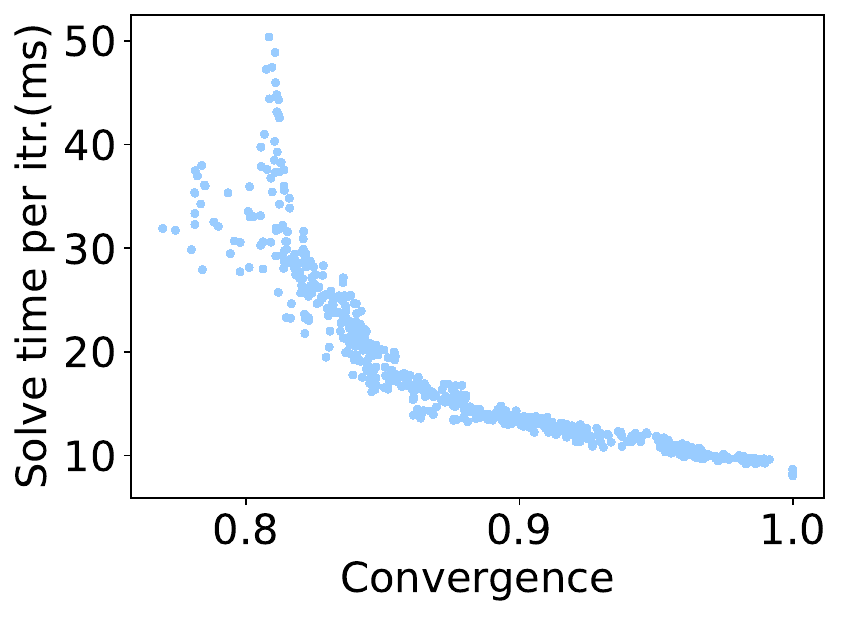} \\
          \bottomrule
    \end{tabular}
    \caption{Evolution of \ac{AMG}-preconditioned GMRES programs (blue dots) with respect to \textit{solve time per iteration} and \textit{convergence}, progressing from an initial random population to the final generation (clockwise from top left: \textit{initial population}, \textit{generation 1}, \textit{generation 10}, \textit{generation 100}).}
    \label{fig:evolution}
  \end{figure}

\begin{figure}
      \centering
      \includegraphics[width=1\linewidth]{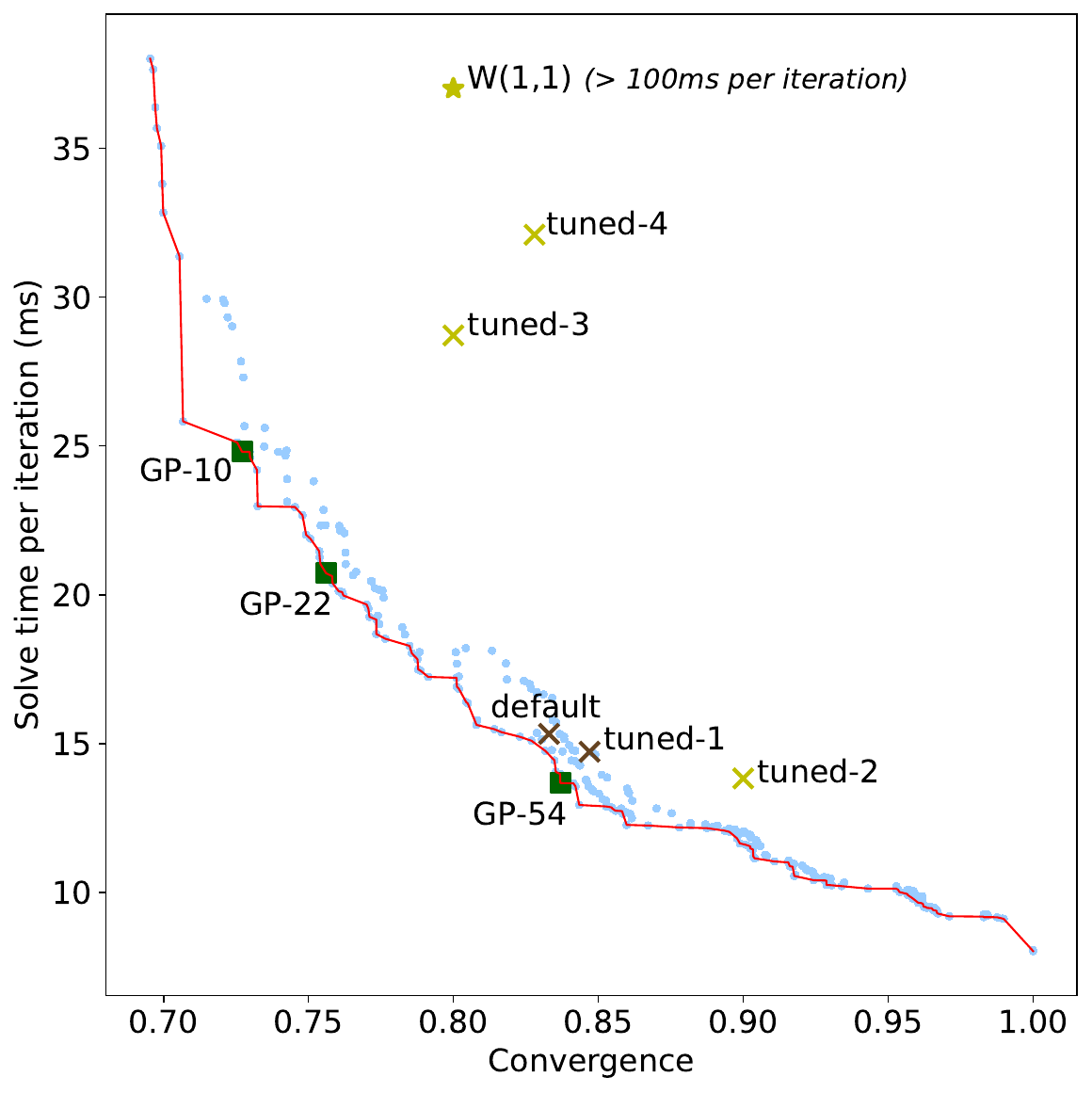}
      \caption{Population of the final 100\textsuperscript{th} generation of AMG evolution. The red line represents the set of Pareto optimal solvers, with green squares indicating the GP solvers selected for our problem. The crosses denote the positions of the reference solvers relative to the Pareto front. The W-cycle is marked differently, as its performance is sub-optimal and the actual position lies outside the figure.}
      \label{fig:paretofront}
  \end{figure}

The evolution of \ac{AMG} preconditioners for the proxy linear system at $t = 0.4\,s$, $2^{nd}$ Newton iteration begins with a random set of solvers generated by the CFG, resulting in many initially diverging solvers. The successive application of genetic operators -- mutation and crossover --  produces solvers with improved convergence and lower cost (\cref{fig:evolution}, clockwise from top-left). As the population evolves over multiple generations, it clusters closer to the Pareto front (red line in \cref{fig:paretofront}).

 We are interested in preconditioners that minimize the \textit{solve time}; and this depends on two factors, the cost for a single iteration and the number of iterations for convergence. Our motivation to split the solve time into two objectives -- \textit{solve time per iteration} and \textit{convergence} -- is to promote diversity in the population, encourage exploration, and prevent the optimization from getting stuck in local minima. However, a fast preconditioner must achieve reasonably minimal values for both objectives. The Pareto optimal solvers from the final generation are ranked by increasing iteration count, indexed, and three solvers—GP-10, GP-22, GP-54—are selected from distinct regions of the Pareto front (represented as green squares in \cref{fig:paretofront}). Solvers on the tail ends of the Pareto front are excluded, as they only minimize one of the two objectives, leaving the remaining options as viable choices for the simulation. The flexible cycle structures for GP-10 and GP-54 are shown in \cref{fig:cycleviz}. The three GP solvers are evaluated and compared to reference methods (see \cref{subsec:refmethods} for reference solver details) across the entire laser beam welding simulation (\cref{subsec:timestepping}) and further tested on different problem sizes (\cref{subsec:weakscaling}).

Specifically, the performance of the GP solvers is compared to the default BoomerAMG configuration provided via the PETSc interface (further referred to as `default’) and a hand-tuned BoomerAMG configuration (denoted as `tuned-1’ in \cref{tab:refsolvers}). The hand-tuning steps are detailed in \cref{subsec:refmethods}, and the position of these reference methods relative to the GP Pareto front is marked in \cref{fig:paretofront}. As seen in \cref{fig:paretofront}, the reference methods (represented as crosses in \cref{fig:paretofront}) are sub-optimal with respect to Pareto optimal GP solvers (for $t=0.4$, $2^{nd}$ Newton iteration). For further evaluation, we select the reference methods — \textit{default} and \textit{tuned-1} — that lie closest to the Pareto front. We observe that the time to setup the different \ac{AMG} preconditioners is the same, since GP and reference solvers use identical \ac{AMG} setup parameters (\cref{tab:amgsetupparam}). Also, as seen in \cref{tab:timings_amg_ref}, the setup times are significantly lower than the solving times. Therefore, in the following sections, we focus solely on the \textit{solve phase}, that is, time spent on solving the linear systems and the corresponding iteration count.
\subsection{Generalization across time steps}\label{subsec:timestepping}
\begin{figure}
    \centering
    \includegraphics[width=1\linewidth]{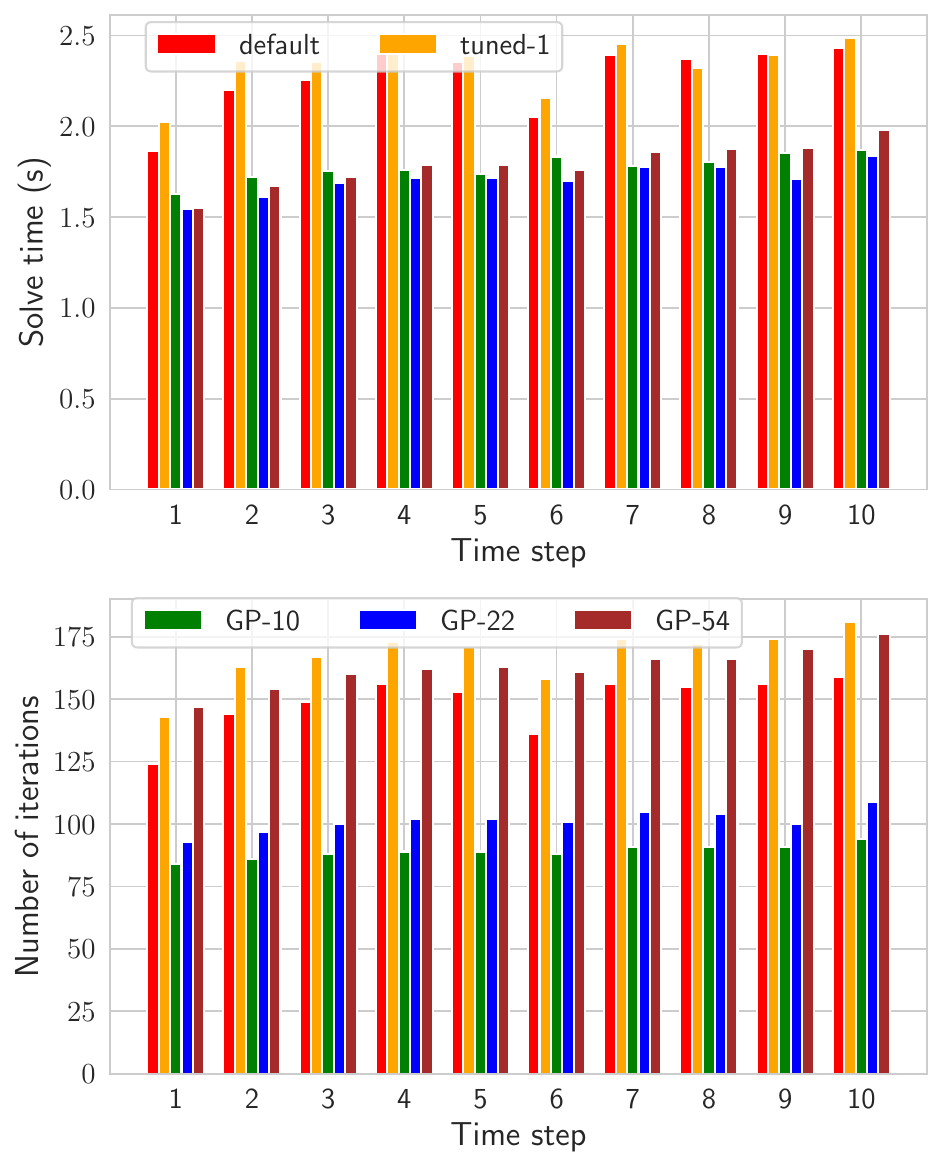}
    \caption{Performance of GP generated \textit{flexible} \ac{AMG} preconditioners and standard \ac{AMG} preconditioners for the numerical simulation with 89\,100 DOFs.}
\label{fig:amgopt_8procs}
\end{figure}

As a first step, we evaluate the GP and reference solvers for $1.0 \, s$ of laser beam welding simulation with \numgru{89100} DOFs distributed across 8 MPI processes. In \cref{fig:amgopt_8procs} we present the solve times and iteration counts at each time step summed over multiple Newton iterations. All GP solvers exhibit lower solve times compared to the reference ones across all time steps, demonstrating the robustness of the GP-generated solvers despite being exposed to just a single linear system of the entire simulation.

The convergence of the GP solvers (\cref{fig:amgopt_8procs}, bottom) is influenced by the region of the Pareto front from which they were selected. GP-10 and GP-22 show the fastest convergence, while GP-54 has a convergence factor similar to that of the reference solvers. As seen in \cref{fig:paretofront}, GP-54 has a lower cost per iteration compared to the reference methods, and therefore achieves a faster time to solution inspite of having a similar convergence rate. This is because GP generates a cost-efficient cycle structure -- GP-54 -- with no pre-smoothing steps on the finest levels and \textit{flexible} post-smoothing sequences (shown in \cref{fig:cycleviz}), leading to cheaper iterations, but without losing convergence.
\subsection{Generalization across problem sizes}\label{subsec:weakscaling}
We now perform a weak scaling study to evaluate solver performance across larger problem sizes. The number of elements per MPI process (tile resolution in \cref{fig:MPIdist}) is fixed. The problem is refined by doubling the number of subdomains (and hence the number of MPI processes) along both plate directions, while keeping the problem resolution orthogonal to the welding plate fixed. Consequently, the total DOFs approximately quadruple at each scaling step, starting from the original problem with \numgru{89100} DOFs. The weak scaling measurements, listed in \cref{tab:weakscaling}, were performed with 8 MPI processes per node on 1, 4, and 16 nodes of the Fritz compute cluster\footnote{https://doc.nhr.fau.de/clusters/fritz/} at the Erlangen National High Performance Computing Center (NHR@FAU).

As shown in \cref{tab:weakscaling} all GP solvers exhibit lower solving times compared to the reference solvers across all problem sizes, with GP-10 and GP-22 demonstrating the fastest convergence rates. $\eta_1$ and $\eta_2$ measure the speedup of the best GP solver (timings in bold) with respect to the default PETSc configuration and hand-tuned BoomerAMG configuration respectively. For the largest problem with 1.37M DOFs, GP-10 is approximately $60\%$ faster than the default BoomerAMG configuration in PETSc and $25\%$ faster than the hand-tuned \ac{AMG} method.

Despite demonstrating that the automated approach allows us to construct efficient and optimal preconditioners for a given \ac{AMG} setup, we observe that the iteration count grows quickly for the largest problem size, resulting in sub-optimal scaling. We observe that the initial Newton iterations exhibit good scaling, while the later iterations create a bottleneck for optimal scaling  (\cref{fig:itrcount_newton}). More details on this are discussed in \cref{subsec:discussion}. Also, treating \ac{AMG} as a black-box and monolithic solver might not be the optimal choice. Instead, variants of block-triangular preconditioners should be used and optimized in a similar automatic fashion. 

\begin{table}[t]
    \centering
\begin{tabular}{lrrrrrrrrrrrrrr}
\toprule
\multicolumn{1}{c}{} & \multicolumn{2}{c}{\textbf{GP-10}} & \multicolumn{2}{c}{\textbf{GP-22}} & \multicolumn{2}{c}{\textbf{GP-54}}\\
\midrule
$DOFs$ &   T(s) &   N &  T(s) &   N &  T(s) &   N \\

\midrule
\numgru{89100}   &     17.7 &  89 &    \textbf{17.1} &  101 &    17.9 &  162  \\ 
\numgru{347116}    &     24.7 &  117 &    \textbf{22.8} &  125 &    24.3 &  206  \\ 
\numgru{1370028}    &     \textbf{277.6} &  984 &    279.1 &  1155 &    326.8 &  2003  \\
\bottomrule
\end{tabular}
\begin{tabular}{lrrrrrrrrrrrrrr}
\toprule
\multicolumn{1}{c}{} & \multicolumn{2}{c}{\textbf{default}} & \multicolumn{2}{c}{\textbf{tuned-1}} & \multicolumn{2}{c}{\textbf{speedup}} \\
\midrule
$DOFs$ &  T(s) &   N &  T(s) &   N & $\eta_1$ & $\eta_2$\\

\midrule
\numgru{89100}    &    22.7 &   149 &    23.3 &  168 &  1.33 &    1.37 \\ 
\numgru{347116}   &    31.9 &   188 &    30.9 &  210 &   1.40 &    1.36 \\ 
\numgru{1370028}   &    436.7 &  1942 &    341.7 &  1665 &   1.57 &    1.23 \\ 
\bottomrule
\end{tabular}
    \caption{Total solving time and average iteration count per time step for different problem sizes.}
    \label{tab:weakscaling}
\end{table}

\section{Conclusions}\label{subsec:discussion}

\begin{figure*}[t]
  \centering
        \includegraphics[width=0.7\linewidth]{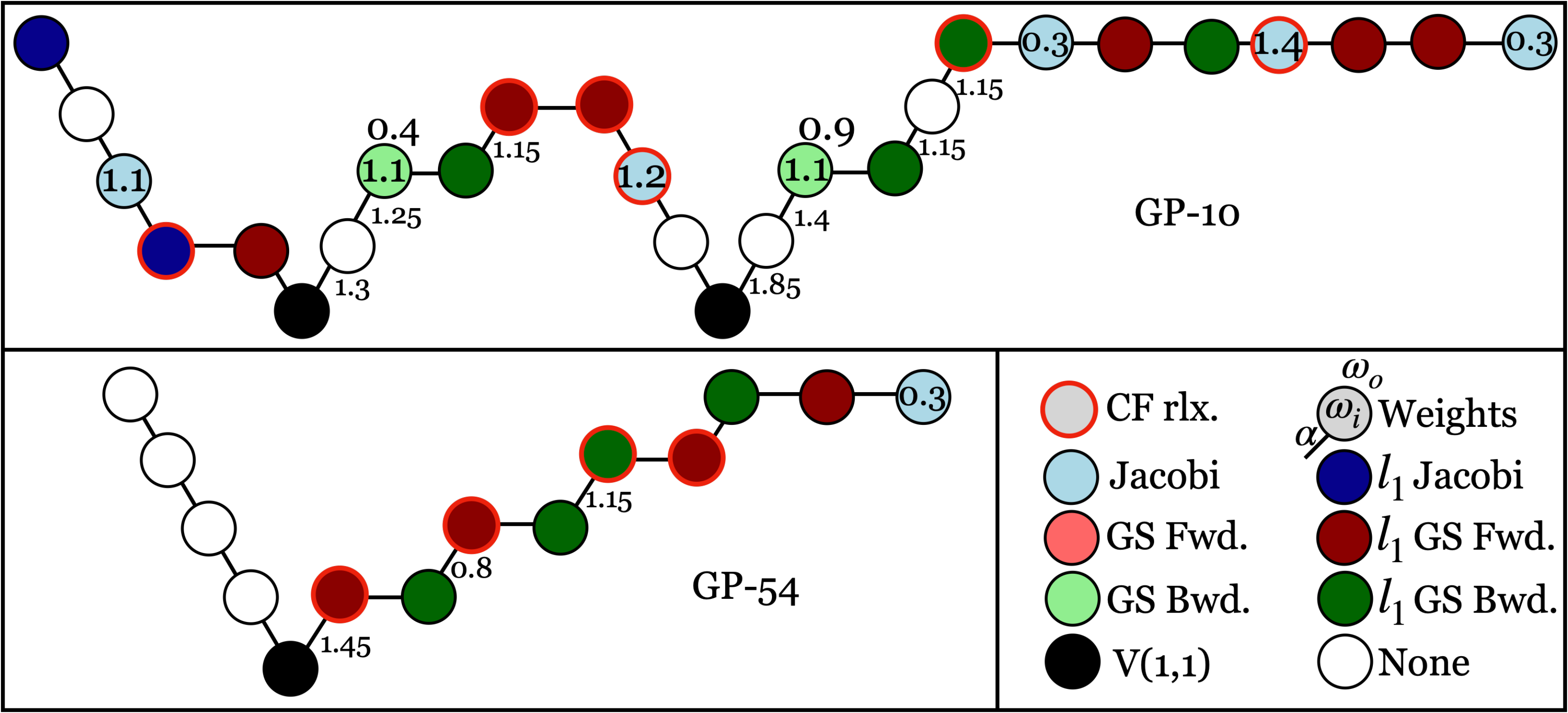}
        \caption{Visual representation of the GP generated flexible cycles.}
        \label{fig:cycleviz}
\end{figure*}

\begin{figure}[b]
    \centering
        \includegraphics[width=\linewidth]{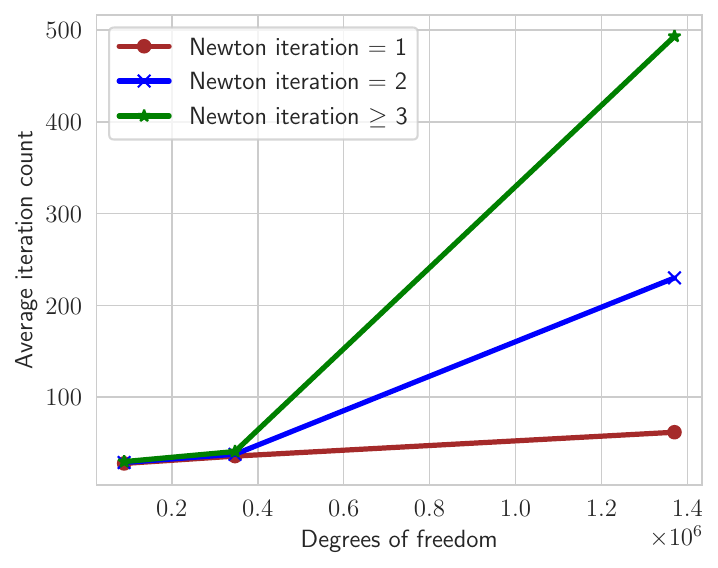}
        \caption{Average iteration count of GP-10 for different Newton iterations.}
        \label{fig:itrcount_newton}
\end{figure}

As previously stated, when we measure the convergence of GP-10 separately for different Newton iterations during the simulation (\cref{fig:itrcount_newton}), the solver scales well for the initial Newton iterations but is sub-optimal for the later steps. This trend, also observed in the reference solvers, could be caused by an algorithmic saturation, that is, fast converging error modes are solved initially, leaving the remaining slower modes for the later steps; or due to poor conditioning of the linear systems across Newton's method.
Our current approach generates solvers using a small proxy problem on a single node and then applies them at a larger scale for the entire laser beam welding simulation. A potential improvement could be to incorporate fitness measures directly from the larger problem, where scaling limitations are observed, as an additional fine-tuning step. For example, the final population of solvers from the small problem, could serve as the initial population for a larger problem, and this can be done progressively to evict solvers with bad scaling from the population. 

We emphasize that in this paper, our goal was to find the best configuration for a monolithic black-box BoomerAMG. By leveraging the advantages of GP, we generate flexible cycles with level-dependent smoothing sequences, cycling patterns, and obtain a variety of optimal preconditioners with different strengths (GP-10 / GP-54, \cref{fig:cycleviz}), based on their position on the Pareto front. These have the following implications:

\begin{itemize}
    \item Possibility to design smoothing sequences tailored to specific right-hand sides, for example, to help mitigate sub-optimal scaling of specific Newton steps.
    \item Access to a set of Pareto optimal solvers provides flexibility in choosing preconditioners of different strengths depending on the conditioning of the linear system.
    \item Since \ac{AMG} cycles are generated for the same setup; once the \ac{AMG} hierarchy is established, multiple flexible cycles (for example, for different Newton iterations) can be used during the simulation without additional overhead.
\end{itemize}

To conclude, in general, the GP approach helped to find a better black-box BoomerAMG for thermo-elastic laser beam welding simulations, but still several improvements can be made. Additionally, more efficient approaches can be studied in the future, like including specific types of smoothers which are more suitable for coupled systems and nodal coarsening approaches which also might interpolate the null-space of the differential operators exactly~\cite{nodal1,nodal2}, into the search space. Also considering block-triangular preconditioners with optimized BoomerAMG for the blocks might be beneficial.


\begin{acks}
This project has received funding from the Deutsche Forschungsgemeinschaft (DFG) as part of the Forschungsgruppe (Research Unit) 5134 ``Solidification Cracks During Laser Beam Welding -- High Performance Computing for High Performance Processing''.
The authors gratefully acknowledge the scientific support and HPC resources provided by the Erlangen National High Performance Computing Center (NHR@FAU) of the Friedrich-Alexander-Universit\"at Erlangen-N\"urnberg (FAU) under the NHR project k109be10. NHR funding is provided by federal and Bavarian state authorities. \\
NHR@FAU hardware is partially funded by the German Research Foundation (DFG) - 440719683. We would like to thank our project partners from Forschungsgruppe 5134 L.\ Scheunemann, J. Schr\"oder, and P. Hartwig for providing the thermo-elasticity formulation, the material parameters and collaborating with us on the FEAP interface, and also M. Rethmeier and A. Gumenyuk for providing the surface data of the melting pool.
\end{acks}


\bibliographystyle{ACM-Reference-Format}
\bibliography{literature}


\appendix
\section{Software details}\label{subsec:software}
We use \textit{EvoStencils}, a software package for automated multigrid design \cite{Schmitt2021,Schmitt2022}, to generate efficient AMG preconditioners for our simulation. This package utilizes the DEAP library \cite{DEAP_JMLR2012} to enforce grammar constraints and formulate \acp{CFG}. \textit{EvoStencils} contains grammar rules for BoomerAMG, incorporating its different parallel smoothers and the corresponding parameter choices for each smoother type \cite{Baker2011}. Thus, it can leverage the existing BoomerAMG solver options, with added cycle flexibility, and generate grammar expressions for AMG individuals. These grammar expressions are transformed into BoomerAMG parameters. Additional interfaces have been added and implemented in a forked version of \textit{hypre} to enable flexible cycling with existing BoomerAMG options \cite{ParthasarathyCM2024}.

\begin{figure}
    \centering
\includegraphics[width=1\linewidth]{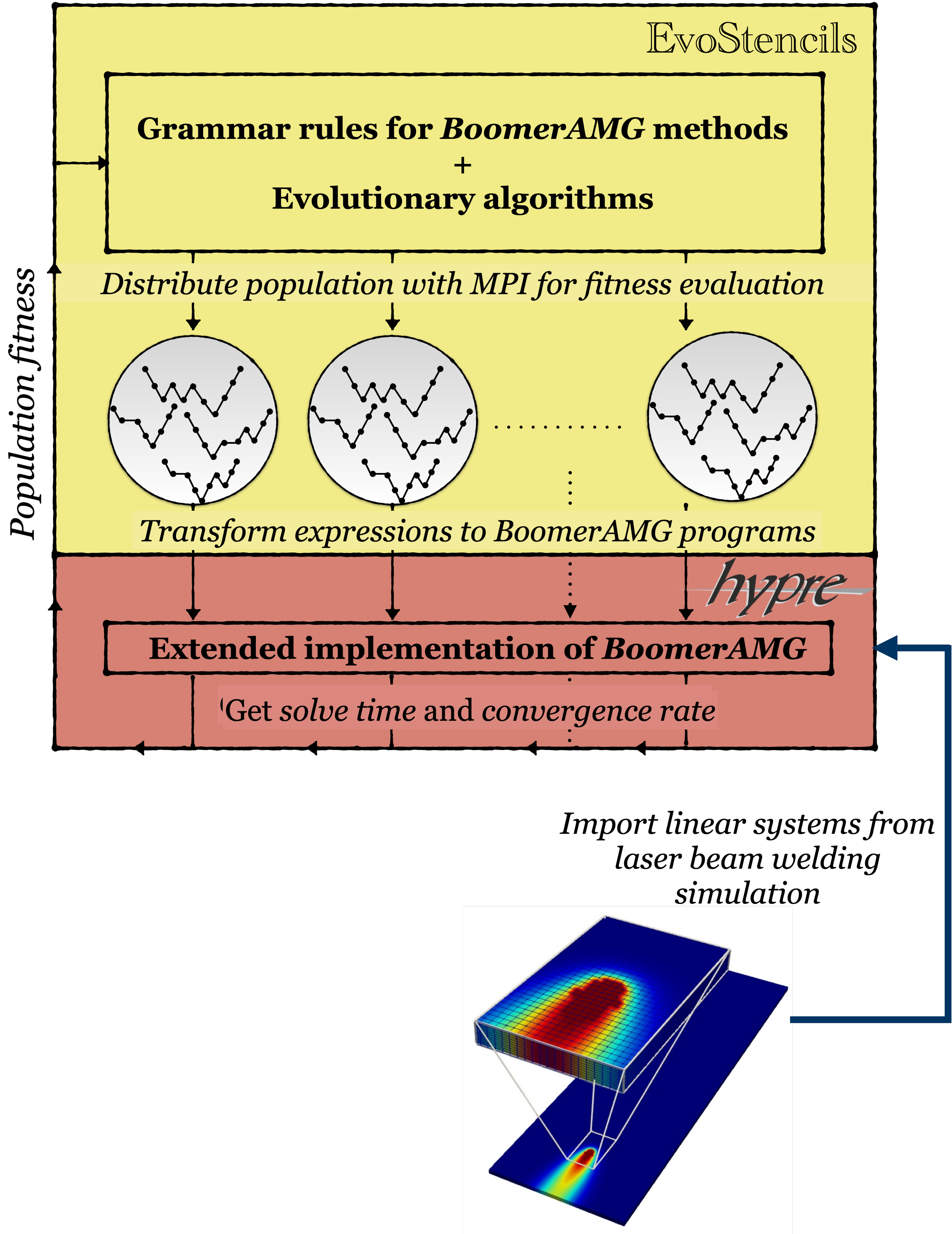}
    \caption{Overview of the software pipeline for automated \ac{AMG} design.}
    \label{fig:softwareoverview}
\end{figure}

\cref{fig:softwareoverview} illustrates the software pipeline for our problem. The CFG for BoomerAMG generates a population of AMG individuals as grammar expressions, which are then split across MPI processes for parallel fitness evaluation. These grammar expressions are transformed into BoomerAMG parameters, and the corresponding programs are executed on the target platform. GMRES preconditioned with these AMG individuals solves the linear system(s) imported from the laser beam welding simulation into \textit{hypre}. The solve times and convergence rates measured are fed back to \textit{EvoStencils} as fitness measures. Crossover and mutation are applied to the grammar representation of selected individuals with high fitness, generating offspring and evolving the next generation. In short, the grammar expressions act as the genotype\footnote{ In biological terms, genotype refers to the genetic makeup or DNA sequence of an individual, while phenotype describes its observable traits, which are determined by both the genetic makeup and environmental influences.}, to which genetic operators—crossover and mutation—are applied. The transformed BoomerAMG programs act as the phenotype, used to measure the quality of each individual. These steps are iterated as shown in \cref{fig:GPoverview} for a fixed number of generations, and finally, we obtain a set of Pareto optimal BoomerAMG programs.

Furthermore, we use the ($\mu + \lambda$) evolutionary strategy for our optimization. This approach combines the current population ($\mu$) with their offspring ($\lambda$), and the combined population is sorted using NSGA-II \cite{NSGA}. For the results presented in the paper, we evolve \ac{AMG} individuals for 100 generations with a population size of 256. Summary of key GP parameter settings are listed in \cref{tab:g3pparam}.

\begin{table}
\centering
\begin{tabular}{|r|l|}
\toprule
        \textbf{Evolutionary algorithm} & ($\mu + \lambda$) \\
        \textbf{Generations}  & 100 \\
        \textbf{$\mu$\,(population)}  & 256 \\
        \textbf{$\lambda$\,(offspring)}  & 256 \\
        \textbf{Initial population factor }  & 64 \\
        \textbf{Crossover probability}  & 0.9 \\
        \textbf{Sorting algorithm}  & NSGA-II \\
        \bottomrule
      \end{tabular}
      \caption{A summary of key GP parameters.}
      \label{tab:g3pparam}
\end{table}
\section{Reference methods}\label{subsec:refmethods}
\begin{table}
\centering
\begin{tabular}{|r|l|}
\toprule
        \textbf{Coarsening} & Falgout coarsening \\
        \textbf{Interpolation}  & Extended+i \\
        \textbf{Strength threshold}  & 0.8 \\
        \textbf{Max. row sum}  & 0.9 \\
        \bottomrule
      \end{tabular}
      \caption{A summary of key \ac{AMG} setup parameters.}
      \label{tab:amgsetupparam}
\end{table}
We integrate the PETSc interface to BoomerAMG into the laser beam welding simulation software and use \ac{AMG}-preconditioned GMRES (one \ac{AMG} cycle per GMRES iteration) to solve the sequence of linear systems. The \ac{AMG} setup parameters are hand-tuned for optimal performance and listed in \cref{tab:amgsetupparam}.

With this \ac{AMG} setup, the default BoomerAMG solve configuration in PETSc is chosen as one of the references. To establish more competitive benchmarks, we explore a larger variety of solver options provided by the \textit{hypre} interface to BoomerAMG (with the same \ac{AMG} setup) and hand-tune the solve parameters for pre-smoothers ($S_{pre}$), post-smoothers ($S_{post}$), number of pre-smoothing steps ($N_{pre}$), number of post-smoothing steps ($N_{post}$), inner relaxation weights ($\omega_i$), and outer relaxation weights ($\omega_o$). Performance for W-cycles was sub-optimal, so only V-cycle configurations were explored.
Some of the high-performing tuned configurations are listed in \cref{tab:refsolvers}, with the most optimal one -- tuned-1 -- chosen as the second reference. 

\begin{table}[ht]
    \centering
\begin{tabular}{llllllll}
\toprule
method &    CF & $S_{pre}$ & $N_{pre}$ & $S_{post}$ & $N_{post}$ & $\omega_i$ & $\omega_o$\\
\midrule
\textbf{default} & true &  GS Sym. &         1 &   GS Sym. &          1 &   1.0 &   1.0  \\
 \textbf{tuned-1} &  false &  GS Sym. &         1 &   GS Sym. &          1 &   1.0 &   1.0  \\
\textbf{tuned-2} &   false &  GS Fwd. &         1 &   GS Bwd. &          1 &   0.8 &   1.0  \\
\textbf{tuned-3} &  false &  GS Sym. &         3 &   GS Sym. &          3 &   1.0 &   1.0  \\
\textbf{tuned-4} &   false &  GS Fwd. &         3 &   GS Bwd. &          3 &   0.8 &   1.0  \\
\bottomrule
\end{tabular}
\caption{V-cycle BoomerAMG parameters for default and tuned reference \ac{AMG} methods.}
    \label{tab:refsolvers}
\end{table}

\end{document}